# $ReS_2$-based field-effect transistors and photodetectors

*Enze Zhang[1,2]†, Yibo Jin[1]†, Xiang Yuan[1,2], Weiyi Wang[1,2], Cheng Zhang[1,2], Lei Tang[1,2], Shanshan Liu[1,2], Peng Zhou[3]*, Weida Hu[4]*and Faxian Xiu[1,2]**

[1]State Key Laboratory of Surface Physics and Department of Physics, Fudan University, Shanghai 200433, China

[2]Collaborative Innovation Center of Advanced Microstructures, Fudan University, Shanghai 200433, China

[3]State Key Laboratory of ASIC and System, Department of Microelectronics, Fudan University, Shanghai 200433, China

[4]National Laboratory for Infrared Physics, Shanghai Institute of Technical Physics, Chinese Academy of Sciences, Shanghai 200083, China

* E-mails: faxian@fudan.edu.cn, wdhu@mail.sitp.ac.cn, pengzhou@fudan.edu.cn.

† These authors contributed equally to this work.

**TOC**

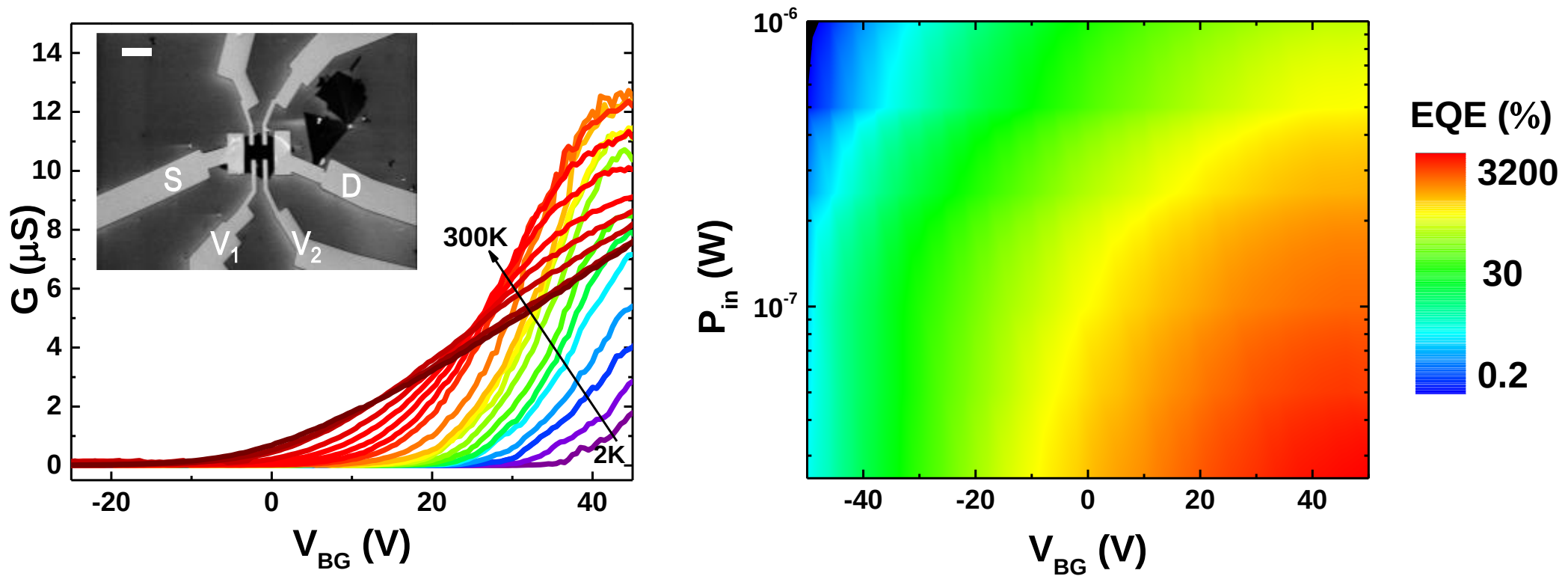

**ABSTRACT:** Atomically-thin two-dimensional (2D) layered transition metal dichalcogenides (TMDs) have been extensively studied in recent years because of their appealing electrical and optical properties. Here, we report on the fabrication of $ReS_2$ field-effect transistors via the encapsulation of $ReS_2$ nanosheets in a high-κ $Al_2O_3$ dielectric environment. Low-temperature transport measurements allowed us to observe a direct metal-to-insulator transition originating from strong electron-electron interactions. Remarkably, the photodetectors based on $ReS_2$ exhibit gate-tunable photoresponsivity up to 16.14 A/W and external quantum efficiency reaching 3,168%, showing a competitive device performance to those reported in graphene, $MoSe_2$, GaS and GaSe-based photodetectors. Our study unambiguously distinguishes $ReS_2$ as a new candidate for future applications in electronics and optoelectronics.

**T**wo-dimensional materials like graphene and $MoS_2$ have been widely studied because of their intriguing physical properties and promising applications in optoelectronic and spintronic devices.[1-4] Due to its mechanical flexibility and high carrier mobility, graphene has been incorporated into transparent devices for electrodes and high-speed photodetectors as channel material.[5-7] Importantly, because of its small intrinsic spin-orbit coupling and vanishing hyperfine interaction, extensive research has been devoted to graphene spintronics.[8, 9] However, despite these substantial efforts, owing to its zero band gap the graphene-channeled devices typically possess a low on/off ratio that significantly impedes its applications.

$MoS_2$, as an exemplary member of two-dimensional (2D) layered transition metal dichalcogenides (TMDs), exhibits a transition from an indirect band gap (1.2 eV) to a direct band gap (1.8 eV) in monolayer, similar to those observed in other TMDs such as $MoSe_2$, $WS_2$ and $WSe_2$.[10,11, 12] $MoS_2$ field-effect transistors (FETs)[13] show a high mobility of 200 $cm^2V^{-1}s^{-1}$ with an on/off current ratio of approximately $10^8$, which can be readily applicable to flash memory,[14-17] high-frequency,[18, 19] valleytronic and spintronic devices.[20-22] However, the exploration of other members in the TMDs family is still at the early stage, and new findings should be anticipated and enlightening.[23, 24]

Among the TMDs family, special attention has been paid to $ReS_2$ because its bulk material behaves electronically and vibrationally as decoupled monolayers arising from the lack of interlayer registry and weak interlayer coupling.[25] Consequently, $ReS_2$ maintains a direct-bandgap of 1.5 eV from bulk to monolayer.[25,

[26] In contrast with what has been discovered in $MoS_2$ and other 2D-TMDs, the unique band structure of $ReS_2$, irrespective to its thickness, offers exciting opportunities for achieving high-efficiency photodetectors. Although $ReS_2$ has been tested as a candidate material for field-effect transistors (FETs),[27] there remains an intense interest in improving the device performance and exploring low-temperature transport properties. Here, we demonstrate the fabrication of top-gate FET devices and show an on/off current ratio exceeding $10^6$ at room temperature. As the measurement temperature is reduced, a metal-to-insulator transition is observed in few-layer $ReS_2$ devices by modulating the electric field, indicating strong electron-electron interactions. Furthermore, the $ReS_2$-based photodetectors provide a gate-tunable photoresponsivity up to 16.14 A/W and a high EQE of 3,168%.

## RESULTS AND DISCUSSION

Few-layer $ReS_2$ nanosheets were synthesized on $SiO_2$/Si substrates in a chemical vapor deposition (CVD) furnace using $ReO_3$ and sulphur as the source materials (see methods). Figure 1a shows a typical atomic structure of $ReS_2$ that crystallizes in a distorted 1T structure with clustering of $Re_4$ units forming a one-dimensional chain inside each monolayer (green dash line).[25, 26, 28] Figure 1b illustrates the photoluminescence spectra taken from 3 layers of $ReS_2$ to the bulk, where the band gap experiences negligible change from 1.54 to 1.50 eV, consistent with former studies.[25, 27] Figure 1c displays the optical images of as-grown $ReS_2$ nanosheets. Different thickness can be verified though the optical contrast and precisely determined by atomic force microscopy (AFM). As expected, Raman spectroscopy shows no dependence on thickness due to decoupling of lattice vibrations between adjacent layers (Figure 1d).[25]

Top-gate FET devices were fabricated by performing the standard *e*-beam lithography (EBL) and metal deposition process (see methods). Figure 2a shows the schematic device structure. A 30-nm-thick $Al_2O_3$ layer was deposited by atomic layer deposition (ALD) and the resultant $Al_2O_3$ serves as the top-gate dielectric material. The corresponding scanning electron microscopy (SEM) picture of the device is displayed in Figure 2c. The Fermi level of the encapsulated $ReS_2$ channel can be effectively tuned by changing the top-gate voltage ($V_{TG}$) or the back-gate voltage ($V_{BG}$) applied on the degenerately-doped silicon substrate. To study the gate modulation of the $ReS_2$ nanosheets, source-drain current-voltage ($I_{DS}$-$V_{DS}$) curves at different $V_{TG}$ were measured (Figure 2b): $I_{DS}$ varies linearly with $V_{DS}$, indicating well-developed Ohmic contacts between the electrodes and the $ReS_2$ channel. The transfer curves of the device ($I_{DS}$-$V_{TG}$) can be obtained by sweeping $V_{TG}$ while keeping the back gate grounded. As illustrated in Figure 2d, a maximal on/off ratio more than $10^6$ is acquired when $V_D$ reaches 500 mV ($I_{DS}$-$V_{BG}$, inset of Figure 2d). Meanwhile, a subthreshold swing of 750 mV per decade is extracted, which is comparable to the reported top-gate $MoS_2$ FET devices.[13] Also, the calculated field-effect mobility of the device is about 1 $cm^2V^{-1}s^{-1}$ using the equation $\mu = (dI_{DS}/dV_{TG})\,[L/(WC_i\,V_{DS})]$, where $L$ and $W$ are the channel length and width, respectively. $C_i$ is the capacitance per unit area between the top gate and the $ReS_2$ channel.

Temperature-dependent transport properties were investigated using back-gated four-terminal devices, as shown in Figure 3b inset. For this type of device, the measured four-probe conductance is defined as $G = I_{DS}/ (V_1\text{-}V_2)$, where $I_{DS}$ is the source drain current and $V_1\text{-}V_2$ is the measured voltage drop between the middle two voltage probes. Figure 3a shows double sweeps of $I_{DS}\text{-}V_{DS}$ characteristics for the temperature range of 2 to 300 K at a constant $V_{BG}$ of 40 V. Negligible hysteresis is observed and nonlinear behavior starts to disappear for the temperature above 200 K, excluding the possible influence of the contact resistance or Schottky barrier on the mobility extraction.[29] The dependence of the channel conductance $G$ on $V_{BG}$ at different temperatures is revealed in Figure 3b. When $V_{BG}$ is larger than 15 V a metallic state associated with the metal-to-insulator transition is evident due to the increase of Fermi level.[30] Also, the temperature-dependent field-effect mobility of the device, calculated by using the field-effect mobility formula, reaches its maximum value at 120 K (Figure 3c). Below this critical temperature, the mobility decreases owing to scattering from the charged impurities. Similar behavior has been observed from $MOS_2$ back-gate devices.[29] On the other hand, an increase in temperature has also led to the strong decrease of the mobility, which can be attributed to the electron-phonon scattering that dominates at high temperatures. The temperature dependence follows the relation $\mu \propto T^{-\gamma}$ and the temperature damping factor γ depends on electron-phonon coupling in the sample. By fitting the curve, $\gamma \sim 2.6$ can be extracted, which is slightly larger than that of $MoSe_2$ and $MoS_2$.[4, 24, 29] It is believed that further suppression of phonon scattering can be realized by employing the suitable substrate and the encapsulation of material in high-κ environment.[31]

Figure 3d unveils the dependence of conductance $G$ on temperature under different $V_{BG}$. A metallic behavior typically occurs at high temperatures. The dashed line suggests that the temperature regime for the metallic behavior is enlarged when $V_{BG}$ becomes larger, which is consistent with the previous studies on $WS_2$ and $MoS_2$[29, 32] Also, below 120 K the variation of $G$ weakens for all $V_{BG}$ values (Figure 3d). This can be attributed to the hopping of carriers through localized states which drives the system into a strongly localized regime as the hopping becomes dominant at lower temperatures.[33] As shown in Figure 3e, in the insulating regime (70~250 K), the temperature variation of $G$ can be modelled with the thermally activated equation[34] $G(T) = G_0 exp^{-E_a/k_B T}$, where $E_a$ is the activation energy, $k_B$ is the Boltzmann constant and $G_0$ is a constant. The extracted $E_a$, which corresponds to the thermal activation of charge carriers at the Fermi energy into the conduction band, becomes smaller as the Fermi level moves towards the conduction band (Figure 3e inset), in agreement with $MoS_2$ and $WS_2$ FET devices.[29, 32] Moreover, at 20~250 K, the 2D variable range hopping model,[33, 35] characterized by the equation $\sigma \propto \exp^{-(T_0/T)^{1/3}}$, provides an excellent description for the electrical transport of the $ReS_2$ nanosheets. The localization length can be determined using the equation $\xi_{loc} = (13.8/k_B D T_0)^{1/2}$, where $D$ is the density of states. Taking $D \sim 4\times10^{12}$ $eV^{-1}cm^{-2}$ as the typical surface density of charge traps at the $SiO_2$ interface,[36, 37] $\xi_{loc}$ is estimated to be ~5 nm at $V_{BG}$=15 V, consistent with what attained in $MoS_2$ and $WS_2$.[32, 33]

Due to its sizeable direct bandgap of ~1.5 eV, $ReS_2$ is anticipated to be a promising candidate for photodetection applications.[25, 26] In view of this, two-terminal back-gate photodetectors based on few-layer $ReS_2$ were fabricated (see methods for details). As schematically depicted in Figure 4a, this type of device was probed using a focused laser beam (633 nm) and an illumination power ranging from 12.5 to 1000 nW. Figure 4b shows the quasi-linear and symmetric $I_{DS}$-$V_{DS}$ plots of the device with and without the laser illumination. Compared to the dark current background (black curve in Figure 4b), a significant increase of the photocurrent at a fixed $V_{DS}$ bias was observed when the device was illuminated showing a strong dependence on the laser power. This can be explained by the increased number of photon-generated carriers. Time-resolved photoresponse behavior of the device was probed by switching the laser on and off. The device exhibits a stable and repeatable response to the laser illumination (Figure 4c). At $V_{BG}$= -10 V and $V_{DS}$ = 50 mV, the device shows an on/off current of ~56 and ~20 nA, giving an on/off ratio of ~2.8.

To further examine the photoresponse of the few-layer $ReS_2$ nanosheets, the $I_{DS}$-$V_{DS}$ characteristics of the device under different illumination conditions were acquired. Figure 4a-g show that the photocurrent $I_{ph}$, defined as $I_{ph}= I_{illuminated}-I_{dark}$, increases as $V_{BG}$ increases. This is because the high $V_{BG}$ can tune the Fermi level closer to the conduction band of $ReS_2$ which makes it easier for the tunneling and thermionic current to overcome the barrier between the channel and electrodes. Moreover, the photocurrent $I_{ph}$ also increases with the increase of $V_{DS}$ owing to the enhancement of carrier drift-velocity and the reduction of the carrier transit time.[38]

External quantum efficiency (EQE) is an important parameter to reflect the ratio of electrons flowing out of the device in response to impinging photons,[23] and is defined as $EQE = I_{ph}/e\phi = (I_{ph}hv)/(qP_{in})$, where $q$ is the electron charge, $\phi$ is the number of incident photon, $h$ is the Planck constant and $v$ is frequency of incident laser. The calculated EQE as a function of incident laser power and $V_{BG}$ is presented in Figure 6a in a 2D map. The lower right corner exhibits the maximum value of EQE reaching 3,168%, corresponding to a high back-gate voltage $V_{BG}$ of 50 V and a low incident laser power $P_{in}$ of 25 nW, thus showing a competitive device performance to those reported in InSe photodetectors.[23] At a fixed $V_{BG}$=50 V (Figure 6b), the EQE reveals a linear decrease with increasing of the incident laser power. The trap states caused by defects or charge impurities in the $ReS_2$ channel and the adsorbents (???) at $ReS_2/SiO_2$ interface account for such a decrease of EQE.[38, 39] An increase in the laser power could create more traps being filled by photo-induced charge carriers, leading to the saturation of the photocurrent and the dropping of EQE.[38, 39] Photoresponsivity is another important parameter of the photodetectors,[40] defined as the ratio of the photocurrent to the incident laser power $R = I_{ph}/P_{in}$. Remarkably, the maximum value of the EQE corresponds to a photoresponsivity of 16.14 A/W, proving the $ReS_2$ to be a valuable material candidate for photodetection applications in addition to other 2D materials.[7, 31, 41, 42]

## CONCLUSION

We performed a comprehensive study of the electrical and optical properties of few-layer $ReS_2$ nanosheets. Top-gate FET devices exhibit an on/off current ratio over $10^6$. Systematic temperature-dependent transport measurements reveal a metal-to-insulator transition in four-terminal devices. Few-layer $ReS_2$ photodetectors offer a good EQE of 3,168% and high photoresponsivity of 16.14 A/W. Our results clearly identify the $ReS_2$ as a promising candidate for electronics and optoelectronics.

## METHODS

**Sample growth:** Few-layer $ReS_2$ devices were synthesized in a horizontal tube furnace equipped with a 1-inch-diameter quartz tube. $SiO_2$(285nm)/Si substrates were loaded into the furnace and placed face-down above a crucible containing 1 mg of $ReO_3$ (⩾99.9% Alfa Aesar). Another crucible containing 60 mg of sulphur (⩾99.9% Alfa Aesar) was located upstream. The tube furnace was first pumped and flushed with argon gas in order to remove air. During the growth cycle, the tube furnace was maintained at 760℃ for 10 minutes. The argon gas was kept at a flow rate of 70 sccm at atmospheric pressure. The system was cooled down to room temperature naturally after the growth cycle was completed.

**Top-gate $ReS_2$ FET device**: The drain-source electrodes of the device were fabricated by EBL using PMMA/MMA bilayer polymer. Cr/Au (5nm/80nm) electrodes were deposited by *e*-beam evaporation. After lift-off, a 1 nm-thick Al layer was deposited and oxidized in the air for 24 hours, acting as a seed layer for subsequent deposition of an $Al_2O_3$ (30 nm) dielectric layer via ALD. During the ALD process, trimethylaluminum reacted with water to produce $Al_2O_3$ at 200 ℃ in a vacuum chamber. Another EBL and metal deposition process was used for the top-gate (Cr/Au 5nm/80nm) electrode fabrication. Room temperature electrical properties of the devices were measured in a probe station using a semiconductor device parameter analyzer (Agilent, B1500A).

**$ReS_2$ back-gate device for low temperature and optical measurements:** An EBL and metal deposition process was used in the device fabrication. The temperature-dependent transport measurements of the four-terminal device were carried out in a Physical Property Measurement System (PPMS) system (Quantum design) using an Agilent 2912. A focused laser with a wavelength of 633 nm was used for measurements of optical properties of the devices.

# FIGURES

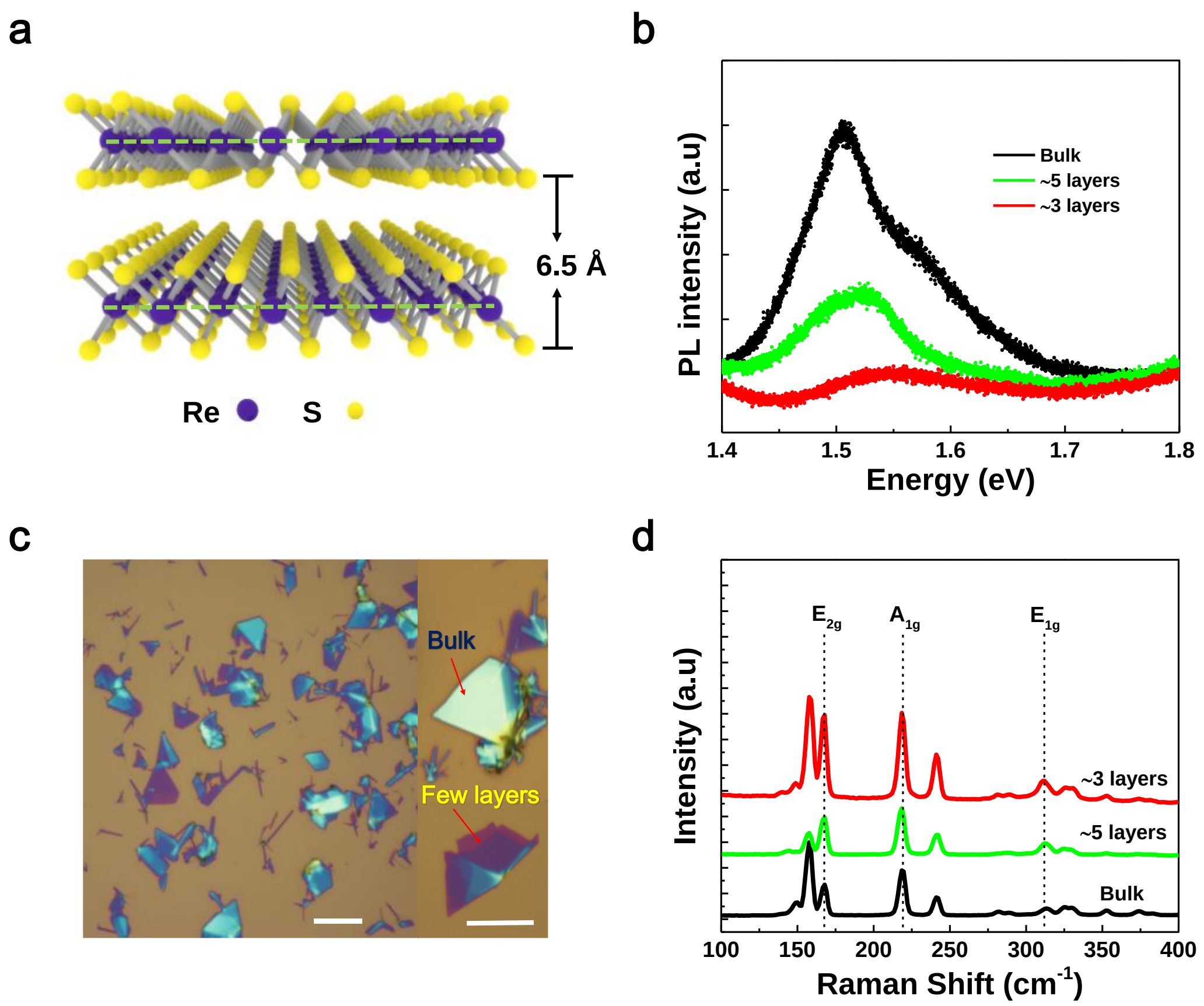

**Figure 1.** Raman and PL spectra of CVD-grown $ReS_2$. (a) Atomic structure of $ReS_2$. The layer to layer distance is ~6.5 Å. (b) Room temperature PL spectra of $ReS_2$ with different thickness. (c) Optical microscopic images of bulk and few-layer $ReS_2$ grown by CVD on $SiO_2$/Si substrate. Scale bars, 5 µm. (d) Raman spectra of $ReS_2$.

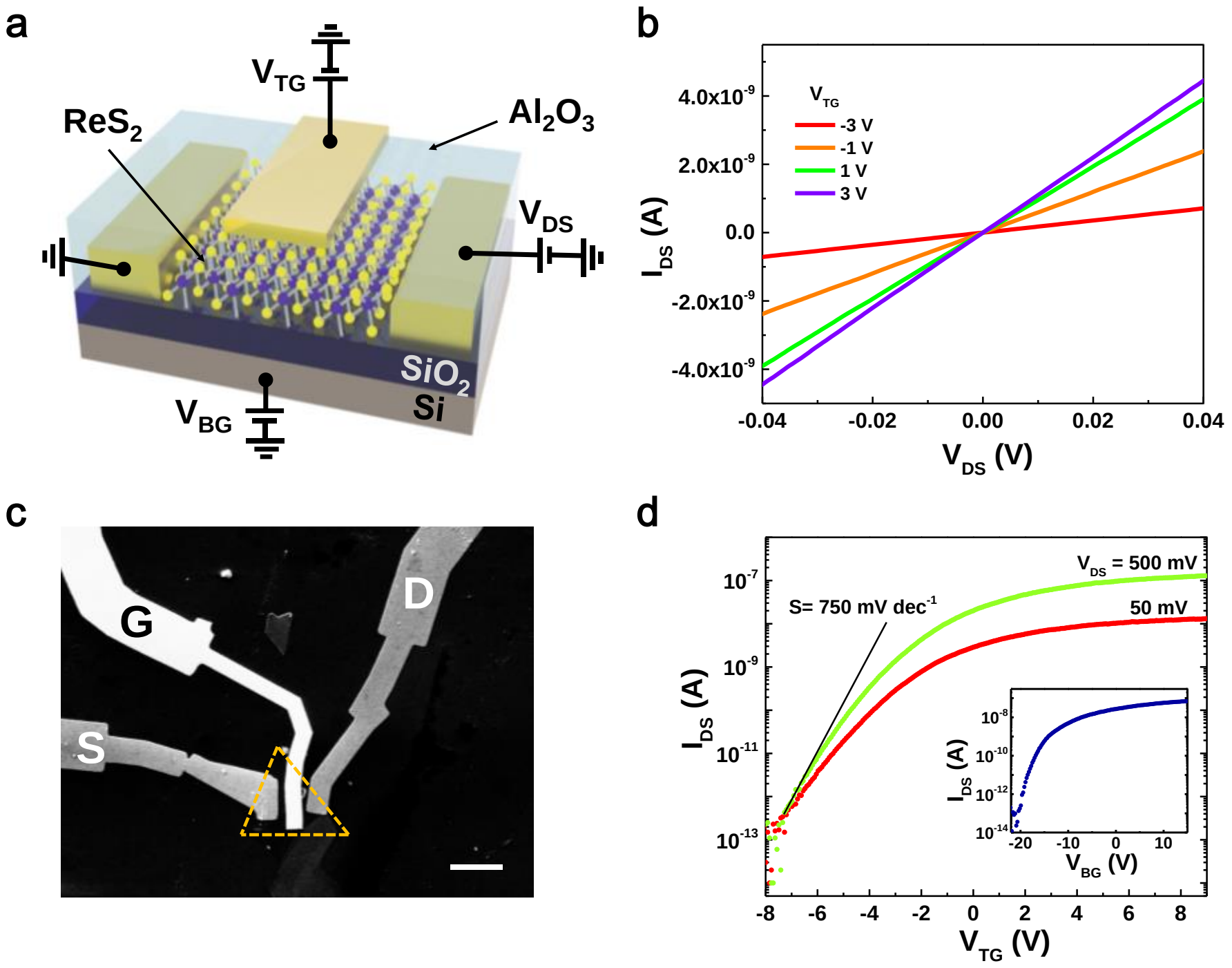

**Figure 2.** Top-gate FET based on few-layer $ReS_2$. (a) Schematic structure of $ReS_2$ top-gate FET. (b) Output characteristics ($I_{DS}$-$V_{DS}$) of the device under different top-gate voltages. (c) SEM image of the fabricated top-gate FET based on few-layer $ReS_2$. Scale bar, 5 μm. (d) Room temperature transfer curves ($I_{DS}$-$V_{TG}$) of the device, which has an on/off current ratio of $10^6$ at $V_{DS}$ = 500 mV. Inset, $I_{DS}$-$V_{BG}$ of the device obtained at $V_{DS}$ = 100 mV.

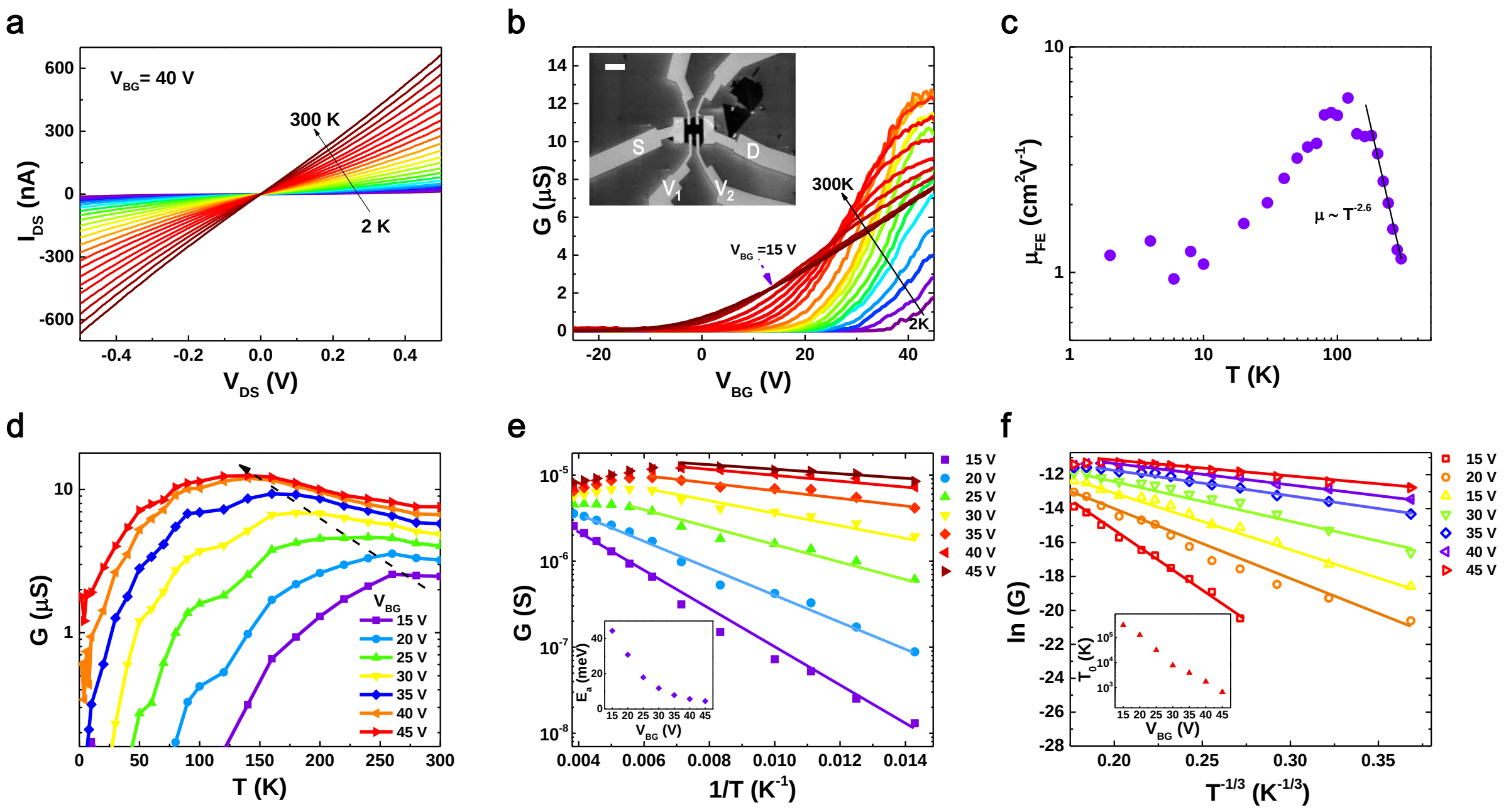

**Figure 3.** Temperature-dependent transport properties of the four-terminal few-layer $ReS_2$ FET. (a) Double sweeps of $I_{DS}$-$V_{DS}$ characteristics for several temperatures at fixed $V_{BG}$ = 30 V with negligible hysteresis. (b) Temperature dependence of the sheet conductivity on back-gate voltage. Inset shows the SEM image of the FET device. Scale bar, 5 μm. (c) Extracted four-terminal field-effect mobility μ vs temperature $T$, The solid black lines are fits to the equation $\mu \propto T^{-\gamma}$ in the temperature range of 120~300 K. (d) Sheet conductivity vs temperature at various back-gate voltages. (e) Arrhenius plots of the sheet conductivity in the high-temperature range (70~250 K). Inset, the activation energy $E_a$ as a function of back-gate voltage $V_{BG}$. (f) Logarithm plots of sheet conductivity as a function of $T^{-1/3}$ for different back-gate voltages, from which the hopping parameter $T_0$ (inset) can be extracted (from the slopes of the line fits).

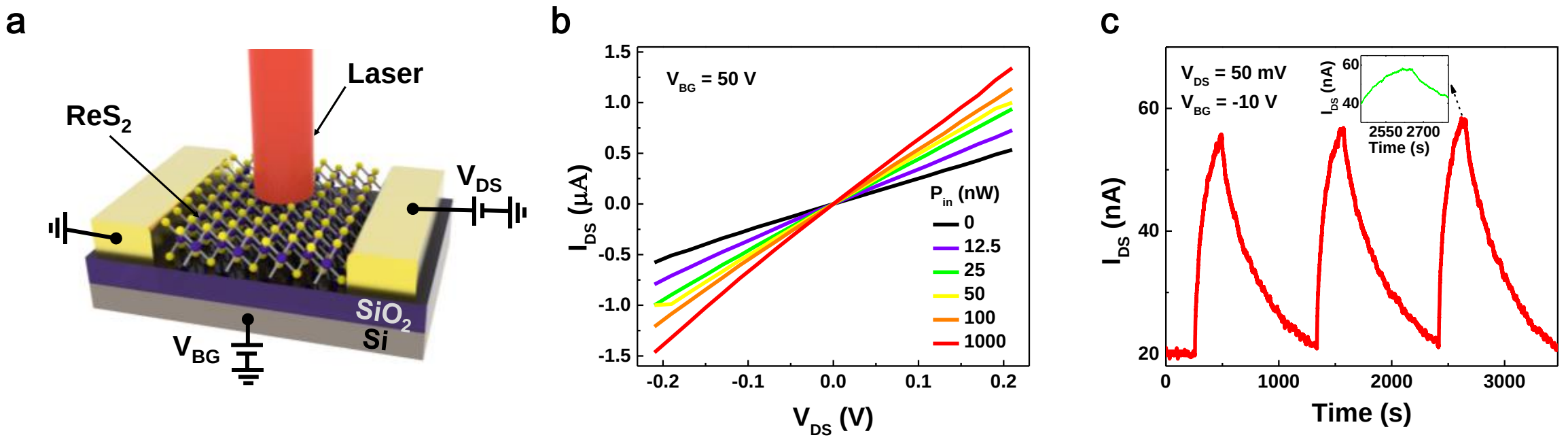

**Figure 4.** Top-gate FET based on few-layer $ReS_2$. (a) Schematic structure of $ReS_2$ back-gate photodetector. (b) Output characteristics ($I_{DS}$-$V_{DS}$) of the device under different incident laser power $P_{in}$. (c) Time-dependent $I_{DS}$ of the device with and without the laser illumination. Inset, plot of the photoresponse peak over a small range, which shows a tendency toward saturation.

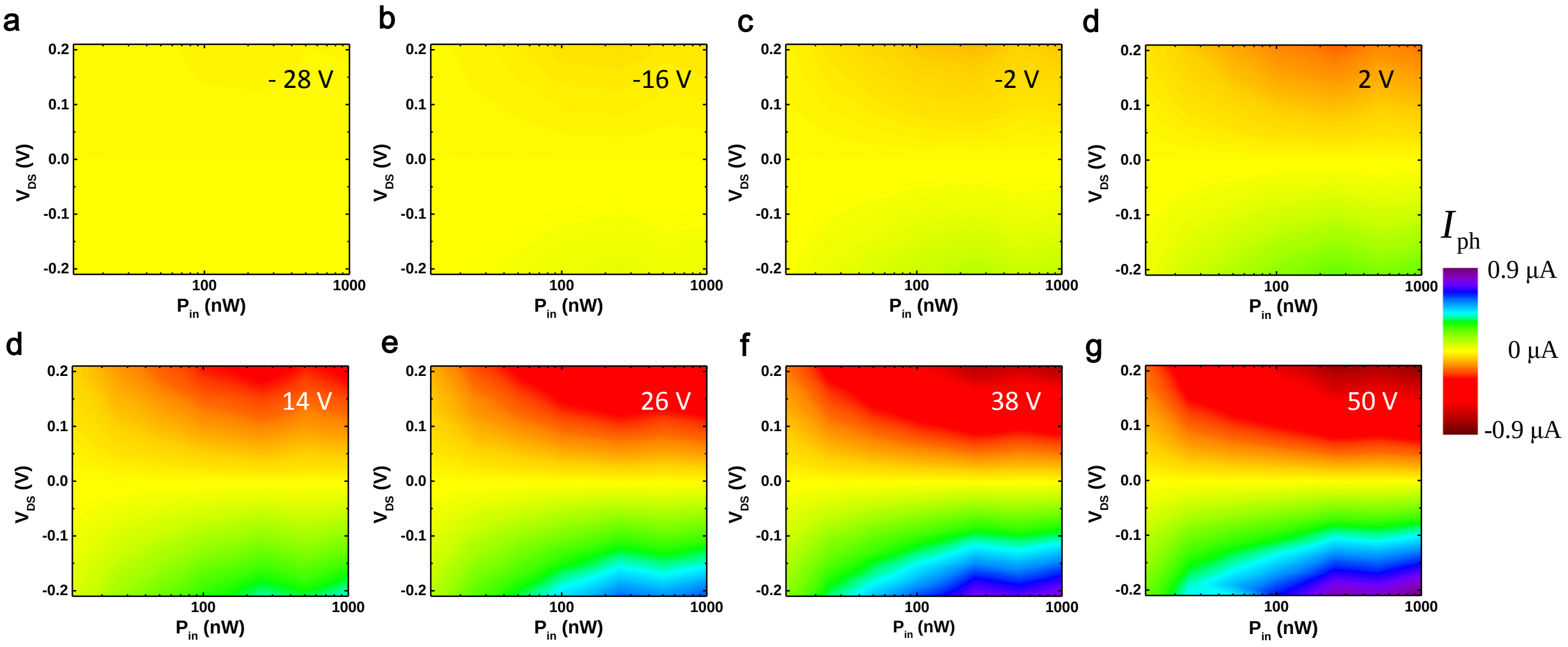

**Figure 5.** The photocurrent $I_{ph}$ versus source-drain voltage $V_{DS}$ and the incident laser power $P_{in}$. (a)-(g) 2D mapping plots of $I_{ph}$ as a function of $V_{DS}$ and $P_{in}$ acquired at different back-gate voltages. The highest $I_{ph}$ corresponds to the largest $V_{DS}$ of 50 V. The increase of $V_{BG}$ gives rise to a higher $I_{ph}$, and a larger $P_{in}$ at certain $V_{BG}$ and also increases $I_{ph}$.

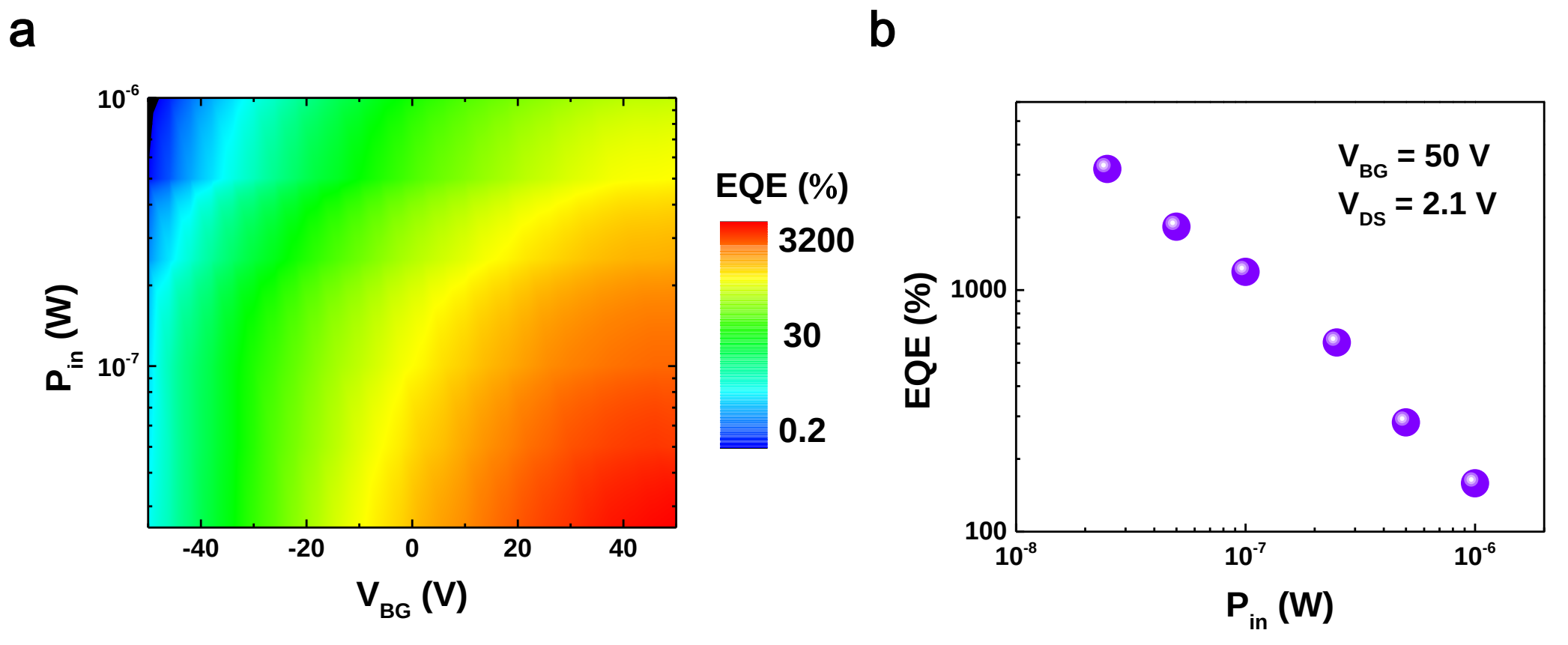

**Figure 6.** Calculated EQE with respect to the incident laser power $P_{in}$ and $V_{BG}$. (a) EQE as a function $P_{in}$ and $V_{BG}$ obtained at $V_{DS}$ = 2.1 V. (b) EQE as a function of $P_{in}$ at fixed $V_{BG}$ = 50 V, where EQE decreases as $P_{in}$ increases due to the enhanced trap states. A high EQE value of 3,168% is obtained when $P_{in}$ = 25 nW.

## REFERENCES AND NOTES

1. Mak, K. F.; McGill, K. L.; Park, J.; McEuen, P. L. The valley Hall effect in MoS2 transistors. *Science* 2014, 344, 1489-1492.
2. Zhang, Y.; Oka, T.; Suzuki, R.; Ye, J.; Iwasa, Y. Electrically switchable chiral light-emitting transistor. *Science* 2014, 344, 725-728.
3. Roy, K.; Padmanabhan, M.; Goswami, S.; Sai, T. P.; Ramalingam, G.; Raghavan, S.; Ghosh, A. Graphene-$MoS_2$ hybrid structures for multifunctional photoresponsive memory devices. *Nat. Nanotechnol.* 2013, 8, 826-830.
4. Chamlagain, B.; Li, Q.; Ghimire, N. J.; Chuang, H.-J.; Perera, M. M.; Tu, H.; Xu, Y.; Pan, M.; Xaio, D.; Yan, J. Mobility Improvement and Temperature Dependence in $MoSe_2$ Field-Effect Transistors on Parylene-C Substrate. *ACS nano* 2014, 8, 5079-5088.
5. Yu, W. J.; Liu, Y.; Zhou, H.; Yin, A.; Li, Z.; Huang, Y.; Duan, X. Highly efficient gate-tunable photocurrent generation in vertical heterostructures of layered materials. *Nat. Nanotechnol.* 2013, 8, 952-958.
6. Mueller, T.; Xia, F.; Avouris, P. Graphene photodetectors for high-speed optical communications. *Nat. Photonics* 2010, 4, 297-301.
7. Xia, F.; Mueller, T.; Lin, Y.-m.; Valdes-Garcia, A.; Avouris, P. Ultrafast graphene photodetector. *Nat. Nanotechnol.* 2009, 4, 839-843.
8. Han, W.; Kawakami, R. K.; Gmitra, M.; Fabian, J. Graphene spintronics. *Nat. Nanotechnol.* 2014, 9, 794-807.
9. Van Tuan, D.; Ortmann, F.; Soriano, D.; Valenzuela, S. O.; Roche, S. Pseudospin-driven spin relaxation mechanism in graphene. *Nat. Phys.* 2014, 10, 857-863.
10. Mak, K. F.; Lee, C.; Hone, J.; Shan, J.; Heinz, T. F. Atomically thin $MoS_2$: a new direct-gap semiconductor. *Phys. Rev. Lett.* 2010, 105, 136805.
11. Tongay, S.; Zhou, J.; Ataca, C.; Lo, K.; Matthews, T. S.; Li, J.; Grossman, J. C.; Wu, J. Thermally driven crossover from indirect toward direct bandgap in 2D semiconductors: $MoSe_2$ versus $MoS_2$. *Nano Lett.* 2012, 12, 5576-5580.
12. Zhao, W.; Ribeiro, R.; Toh, M.; Carvalho, A.; Kloc, C.; Castro Neto, A.; Eda, G. Origin of indirect optical transitions in few-layer $MoS_2$, $WS_2$, and $WSe_2$. *Nano Lett.* 2013, 13, 5627-5634.
13. Radisavljevic, B.; Radenovic, A.; Brivio, J.; Giacometti, V.; Kis, A. Single-layer $MoS_2$ transistors. *Nat. Nanotechnol.* 2011, 6, 147-150.
14. Bertolazzi, S.; Krasnozhon, D.; Kis, A. Nonvolatile memory cells based on $MoS_2$/graphene heterostructures. *ACS nano* 2013, 7, 3246-3252.
15. Wang, J.; Zou, X.; Xiao, X.; Xu, L.; Wang, C.; Jiang, C.; Ho, J. C.; Wang, T.; Li, J.; Liao, L. Floating Gate Memory - based Monolayer $MoS_2$ Transistor with Metal Nanocrystals Embedded in the Gate Dielectrics. *Small* 2015, 11, 208-213.
16. Yin, Z.; Zeng, Z.; Liu, J.; He, Q.; Chen, P.; Zhang, H. Memory Devices Using a Mixture of $MoS_2$ and Graphene Oxide as the Active Layer. *Small* 2013, 9, 727-731.
17. Zhang, E.; Wang, W.; Zhang, C.; Jin, Y.; Zhu, G.; Sun, Q.-Q.; Zhang, D. W.; Zhou, P.; Xiu, F. Tunable charge-trap memory based on few-layer $MoS_2$. *ACS nano* 2015, 9, 612-619.
18. Krasnozhon, D.; Lembke, D.; Nyffeler, C.; Leblebici, Y.; Kis, A. MoS2 transistors operating at gigahertz frequencies. *Nano Lett.* 2014, 14, 5905-5911.
19. Cheng, R.; Jiang, S.; Chen, Y.; Liu, Y.; Weiss, N.; Cheng, H.-C.; Wu, H.; Huang, Y.; Duan, X. Few-layer

molybdenum disulfide transistors and circuits for high-speed flexible electronics. *Nat. Commun.* 2014, 5.

20. Ganatra, R.; Zhang, Q. Few-layer $MoS_2$: A promising layered semiconductor. *ACS nano* 2014, 8, 4074-4099.
21. Mak, K. F.; He, K.; Shan, J.; Heinz, T. F. Control of valley polarization in monolayer $MoS_2$ by optical helicity. *Nat. Nanotechnol.* 2012, 7, 494-498.
22. Zeng, H.; Dai, J.; Yao, W.; Xiao, D.; Cui, X. Valley polarization in MoS2 monolayers by optical pumping. *Nat. Nanotechnol.* 2012, 7, 490-493.
23. Tamalampudi, S. R.; Lu, Y.-Y.; Kumar U, R.; Sankar, R.; Liao, C.-D.; Moorthy B, K.; Cheng, C.-H.; Chou, F. C.; Chen, Y.-T. High Performance and Bendable Few-Layered InSe Photodetectors with Broad Spectral Response. *Nano Lett.* 2014, 14, 2800-2806.
24. Wang, Q. H.; Kalantar-Zadeh, K.; Kis, A.; Coleman, J. N.; Strano, M. S. Electronics and optoelectronics of two-dimensional transition metal dichalcogenides. *Nat. Nanotechnol.* 2012, 7, 699-712.
25. Tongay, S.; Sahin, H.; Ko, C.; Luce, A.; Fan, W.; Liu, K.; Zhou, J.; Huang, Y.-S.; Ho, C.-H.; Yan, J. Monolayer behaviour in bulk $ReS_2$ due to electronic and vibrational decoupling. *Nat. Commun.* 2014, 5.
26. Horzum, S.; Çakır, D.; Suh, J.; Tongay, S.; Huang, Y.-S.; Ho, C.-H.; Wu, J.; Sahin, H.; Peeters, F. Formation and stability of point defects in monolayer rhenium disulfide. *Phys. Rev. B* 2014, 89, 155433.
27. Corbet, C. M.; McClellan, C.; Rai, A.; Sonde, S. S.; Tutuc, E.; Banerjee, S. K. Field Effect Transistors with Current Saturation and Voltage Gain in Ultrathin $ReS_2$. *ACS nano* 2015, 9, 363-370.
28. Murray, H.; Kelty, S.; Chianelli, R.; Day, C. Structure of rhenium disulfide. *Inorganic Chemistry* 1994, 33, 4418-4420.
29. Radisavljevic, B.; Kis, A. Mobility engineering and a metal-insulator transition in monolayer $MoS_2$. *Nat. Mater.* 2013, 12, 815-820.
30. Schmidt, H.; Wang, S.; Chu, L.; Toh, M.; Kumar, R.; Zhao, W.; Castro Neto, A. H.; Martin, J.; Adam, S.; Özyilmaz, B. Transport properties of monolayer $MoS_2$ grown by chemical vapor deposition. *Nano Lett.* 2014, 14, 1909-1913.
31. Jena, D.; Konar, A. Enhancement of carrier mobility in semiconductor nanostructures by dielectric engineering. *Phys. Rev. Lett.* 2007, 98, 136805.
32. Ovchinnikov, D.; Allain, A.; Huang, Y.-S.; Dumcenco, D.; Kis, A. Electrical Transport Properties of Single-Layer $WS_2$. *ACS nano* 2014, 8, 8174-8181.
33. Ghatak, S.; Pal, A. N.; Ghosh, A. Nature of electronic states in atomically thin $MoS_2$ field-effect transistors. *Acs Nano* 2011, 5, 7707-7712.
34. Pradhan, N. R.; Rhodes, D.; Feng, S.; Xin, Y.; Memaran, S.; Moon, B.-H.; Terrones, H.; Terrones, M.; Balicas, L. Field-Effect Transistors Based on Few-Layered α-$MoTe_2$. *ACS nano* 2014, 8, 5911-5920.
35. Paasch, G.; Lindner, T.; Scheinert, S. Variable range hopping as possible origin of a universal relation between conductivity and mobility in disordered organic semiconductors. *Synthetic metals* 2002, 132, 97-104.
36. Ayari, A.; Cobas, E.; Ogundadegbe, O.; Fuhrer, M. S. Realization and electrical characterization of ultrathin crystals of layered transition-metal dichalcogenides. *Journal of applied physics* 2007, 101, 014507-014507-5.
37. Jayaraman, R.; Sodini, C. G. A 1/f noise technique to extract the oxide trap density near the

conduction band edge of silicon. *IEEE Trans. Electron Dev.* 1989, 36, 1773-1782.

38. Lopez-Sanchez, O.; Lembke, D.; Kayci, M.; Radenovic, A.; Kis, A. Ultrasensitive photodetectors based on monolayer $MoS_2$. *Nat. Nanotechnol.* 2013, 8, 497-501.
39. Zhang, W.; Huang, J. K.; Chen, C. H.; Chang, Y. H.; Cheng, Y. J.; Li, L. J. High‐Gain Phototransistors Based on a CVD $MoS_2$ Monolayer. *Adv. Mater.* 2013, 25, 3456-3461.
40. Engel, M.; Steiner, M.; Avouris, P. Black Phosphorus Photodetector for Multispectral, High-Resolution Imaging. *Nano Lett.* 2014, 14, 6414-6417.
41. Hu, P.; Wen, Z.; Wang, L.; Tan, P.; Xiao, K. Synthesis of few-layer GaSe nanosheets for high performance photodetectors. *ACS nano* 2012, 6, 5988-5994.
42. Hu, P.; Wang, L.; Yoon, M.; Zhang, J.; Feng, W.; Wang, X.; Wen, Z.; Idrobo, J. C.; Miyamoto, Y.; Geohegan, D. B. Highly responsive ultrathin GaS nanosheet photodetectors on rigid and flexible substrates. *Nano Lett.* 2013, 13, 1649-1654.